\documentclass[11pt,a4paper]{article}
\usepackage[hyperref]{acl2020}
\usepackage{times}
\usepackage{latexsym}
\usepackage{multicol}
\usepackage{graphicx}
\usepackage{amsmath}
\usepackage{a4wide}
\usepackage[colorinlistoftodos]{todonotes}
 \usepackage{lipsum,tabularx,ragged2e,booktabs}
\usepackage{microtype}

\aclfinalcopy 

\newcommand\MyBox[2]{
  \fbox{\lower0.75cm
    \vbox to 1.7cm{\vfil
      \hbox to 1.7cm{\hfil\parbox{1.4cm}{#1\\#2}\hfil}
      \vfil}%
  }%
}
\title{Using Angle of Arrival for Improving Indoor Localization}

\author{
Sai Koppula \\
  Department of Computer Science\\
  University of Texas at Austin \\
  \texttt{vkoppula@cs.utexas.edu} \\\And
  Shivang Singh \\
  Department of Computer Science\\
  University of Texas at Austin \\
  \texttt{ shivang.singh@utexas.edu} \\}

\begin{document}
\maketitle
\begin{abstract}
    In this paper, we primarily explore the improvement of single stream audio systems using Angle of Arrival calculations in both simulation and real life gathered data. We wanted to learn how to discern the direction of an audio source from gathered signal data to ultimately incorporate into a multi modal security system. We focused on the MUSIC algorithm for the estimation of the angle of arrival but briefly experimented with other techniques such as Bartlett and Capo. We were able to implement our own MUSIC algorithm on stimulated data from Cornell. In addition, we demonstrated how we are able to calculate the angle of arrival over time in a real life scene. Finally, we are able to detect the direction of arrival for two separate and simultaneous audio sources in a real life scene. Eventually, we could incorporate this tracking into a multi modal system combined with video. Overall, we are able to produce compelling results for angle of arrival calculations that could be the stepping stones for a better system to detect events in a scene.
\end{abstract}

\section{Introduction}
Security systems in modern day systems typically rely on a single source type to determine an action. For example, the Ring doorbell system uses a video feed but no audio. The Simplisafe Glass Break detector uses a simple audio feed that is refined to identify the unique sound of a window breaking. We hypothesize that a combination of common monitoring mediums, such as audio, video, and RF, in a multi-modal system can better detect a variety of intrusion types. To this end, we will improve the single stream audio system by using various AoA estimation techniques to localize the source of audio. We implemented the MUSIC algorithm on simulated data to identify the direction of arrival of multiple sources at once. Furthermore, we tried to use the eigenvalues to determine the number of sources present in a simulated environment. Finally, we used the library MUSIC function on captured audio data to track the change of source angles over time. We also explored how we can use machine learning models to further refine our audio system to properly weight the contributions of each of the data sources. Our final product was a working MUSIC implementation and improved audio system that can detect the angles of multiple audio sources. We did not quite integrate the improved audio system into our deficient visual system during the scope of this project.

\section{Related Works}

There are some previous works in localization and anomaly detection. Other works in the field of indoor localization use triangulation techniques to isolate the location of sound. One recent paper, Improving AoA Localization Accuracy, did indoor localization using 5 microphones. They calculated a probability density for each x,y coordinate in the room that they used for their experiment. Our work tries to do something similar with less intrusions to the indoor setup (and having only 1 access point for determining the angle). Furthermore, our work proposes to use indoor cameras to determine the depth of the object. This sacrifices some of the accuracy, but makes the installation and adoption of the system easier to do. \cite{Thoen2019ImprovingAL}. Another paper in this domain is Three-Dimensional Empirical AoA Localization Technique for Indoor Applications. This paper uses factors other than angle of arrival in addition to angle of arrival to do localization. Our work does something similar to this work. However, our work also examines how to do angle of arrival for more than one source. The evaluation in this paper doesn't look at the multipath case. Additionally, the paper looks at RFID signals, while we look at acoustic signals, which are more ubiquitous but also prone to more noise.  \cite{Almaaitah2019ThreeDimensionalEA}. Another paper looks at doing acoustic based angle of arrival estimation for robotics. This paper is similar to our paper. However, there is no distinction made for the multipath case. They do examine the error in the calculation of the real angles versus the projected angle to do evaluation. \cite{Karanam2018MagnitudeBasedAE}


\section{Background}
The core principal we used throughout our project was calculating the angle of arrival (AoA) of a sound source to determine which direction the audio signal was coming from. We envisioned we could project a visualization onto an accompanying video feed to illustrate where the sound was originating from. For example, in an instance when the subject is just waving their hand, the video feed would show changes but the projected audio visualization would not. However, when the subject is snapping their fingers, both the audio and video visualizations would show activity. By checking for overlapping audio and video disturbances, we hypothesized that we could create a better security system than one that just relies on disjoint audio and video feeds. One example of where a multi modal system would excel is where you have multiple sounds in the scene at once. If multiple audio sources are present, an audio only system will tend to alert the user continuously once the audio is past a threshold. However, by mapping the audio onto the scene and correlating it with the various visual disturbances, we can create a more selective system.
\subsection{AoA}
To understand how to achieve an audio visualization, we need to first calculate the angle of arrival. By using an array of antenna elements, we can gather enough data to determine the direction of an audio source. For the same source, various antennas in an array will receive slightly phased version of similar waves. By constructing a linear and equally spaced array of antennas, we can use the expected phase perturbation to correlate what we expect the angle that would have caused such a set of received signals to be. We can accomplish this by leveraging the fact that we can decompose a received signal x as a linear composition of a steering vector S, an amplitude a, and some noise n.  
\begin{equation}
    x = Sa + n
\end{equation}
If we can estimate the steering vector, a function of the angle of the signal hitting an antenna element, we can estimate the angle of arrival. 
\subsection{MUSIC}
The music algorithm leverages the fact that we can decompose the set of linear equations that arise from Equation (1) into their eigenvalues and eigenvectors. The lowest values of these eigenvalues correspond to the noise eigenvectors. It is also known that the noise eigenvectors are orthogonal to the signal sources. We can therefore compose the two across a sweep of angles to judge the orthogonality of the resulting intersection. If the values are close to 0 for a given value of theta, then we must have a good estimate for the angle of arrival. Typically, we take the inverse of this composition to create a pseudospectrum graph with peaks at each sources angles of arrival. \cite{Dave2013DOA}
\subsection{Finding the Number of Sources}
The Music algorithm does have a limitation that it needs to know in advance the number of audio sources. This is so we can properly divide the noise eigenvalues from the signal eigenvalues. However, it is observed that the noise eigenvalues are often quite close to each other while the signal eigenvalues diverge significantly from the noise eigenvalues. Various techniques can be used to dynamically predict the number of sources from analyzing the clustering of eigenvalues\cite{Dave2013DOA}. We explored one such approach in our simulated experiments, but we did not succeed.

\section{Single Stream Visual System}
\subsection{Summary}
In order to determine how multimodal input can help improve the performance of the wireless input, the results and the performance of models for audio and visual detection were studied in isolation. We explored two different approaches for our single stream systems. First, we did a real time stream analysis for a visual system to discern if there is a change in the observable scene.
\subsection{Results}
For the purposes of the single stream video system, we originally planned to utilize a remote Arduino chip, wired with a camera module. This Arduino set up would have a wireless component which would transmit audio and visual data to a server setup. However, the latency and throughput of this configuration was quite poor. For the sake of getting meaningful results, we transitioned to using the webcam on our computer. Here, we were able to stream our computer's webcam into a python application and manipulate the data as needed to determine whether an event occurred. Our algorithm was fairly simple. We kept track of a small history of frames and averaged them together to create a composite frame. We then compared the current frame to the composite frame to see if it differed enough to flag an event. There are a few tunable knobs that can help calibrate the system such as number of frames that contribute to the composite image, the tolerance of a pixel to be different between the composite frame and current frame and the percentage of pixels that can be above that tolerance before an event is flagged. You can see from Figure 1 that our composite image is quite noisy. 

\begin{figure}[htp]
\includegraphics[width=.45\textwidth]{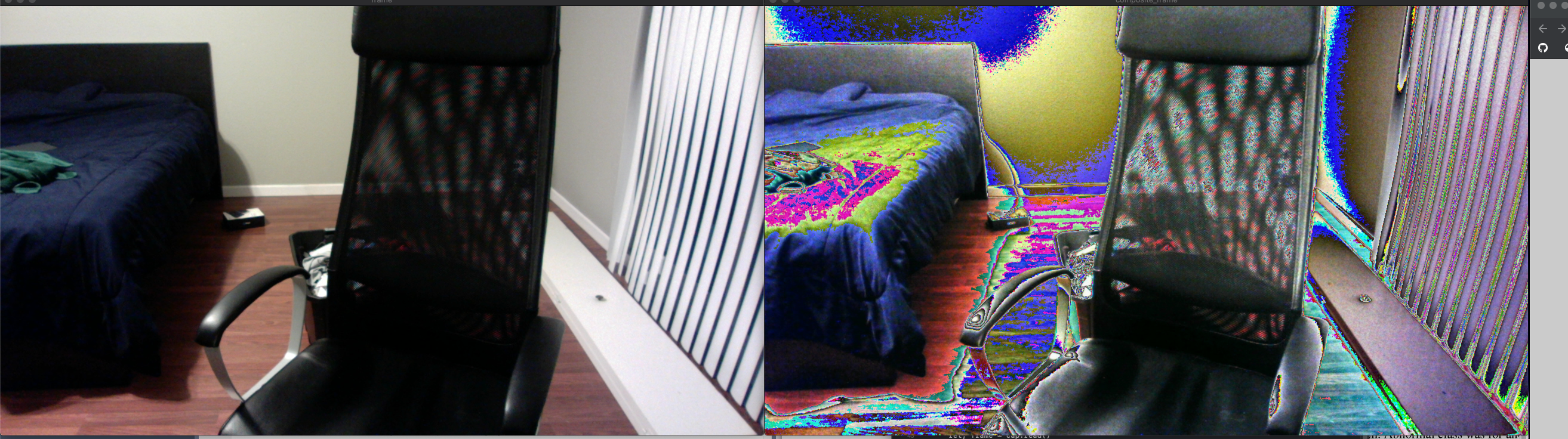}
\label{fig:figure3}
\caption{ Left is the real time camera input. Right is the composite image of the past 5 frames. As you can see, the composite frame is quite noisy and error prone.}
\end{figure}

Overall, our stand alone video system showed quite a few deficiencies. First, it was a noisy image that caused gradients to be present even when nothing in the scene was changing. The noisy image above shows areas where the picture is changing despite nothing occurring in the scene. Second, the visual system suffers from a limited field of view. Any intruder would need to simply avoid the field of view of the camera to avoid detection. Finally, system was sensitive to how it was calibrated. We could get a high number of false positives due to the noisy image and calculation metrics we chose. Whether we chose to take the number of pixels that changed or the average color of the image as a whole, there wasn't a reliable metric to make a robust system from just a video stream. Furthermore, the system could only detect changes but it could not differentiate between different types of changes. In fairness, the audio system by itself, without any machine learning enhancements, suffers from this as well.

\section{Audio System}
The audio system was the main focus of our project. Here we explored how to refine an audio system using AoA techniques including MUSIC to get a better sense of what was happening in the scene over time. We first refined our techniques on simulated data and then moved onto our actual physical setup.We used various python libraries to read our .wav files and transform them into numpy arrays that could be manipulated and calculated on. This results in our current implementation being unable to handle audio in real time. While the algorithm's themselves are real time, we need a post processing step that is not yet automated to bring the data into a manipulative form. 
\subsection{Simulation Setup}
Our simulation setup used data harvested from Cornell University. Here we procured data simulated on a system of a linear array of antennas that then recorded a varying number of signals. We were able to implement our own music algorithm and plot the data against a sweep of angles to find the angle of arrival.

\subsection{Simulation Results}
For implementing the MUSIC algorithm, we did the standard approach of first creating a correlation matrix and then partitioning the noise eigenvalues. We then created pseudospectrum sweeps of the possible angles on the inverse of the orthogonality of the signals and noise eigenvectors. The results are plots below which show the angles from where the sources are from. We tried our algorithm on simulation data for both three sources and two sources. The received signals were from eight antennas in a linear equally spaced array. From the plots below you can see our algorithm partially works for 3 sources. We have peaks near the ground truth values of 18, 54, and 126 degrees. However, we also have extraneous peaks above our 3 source angles which suggests our implementation is not robust for 3 sources. We got much better results for the two source set of simulation data. Here we have two clear peaks near the ground truth values of 54 and 126 degrees. This demonstrated that our MUSIC algorithm was working for at least two sources as we were able to get no extraneous peaks above the ground truth peaks. The results of our experiments from simulation data were good first steps in understanding the MUSIC algorithm for AoA calculation.
\begin{figure}[htp]
\includegraphics[width=.45\textwidth]{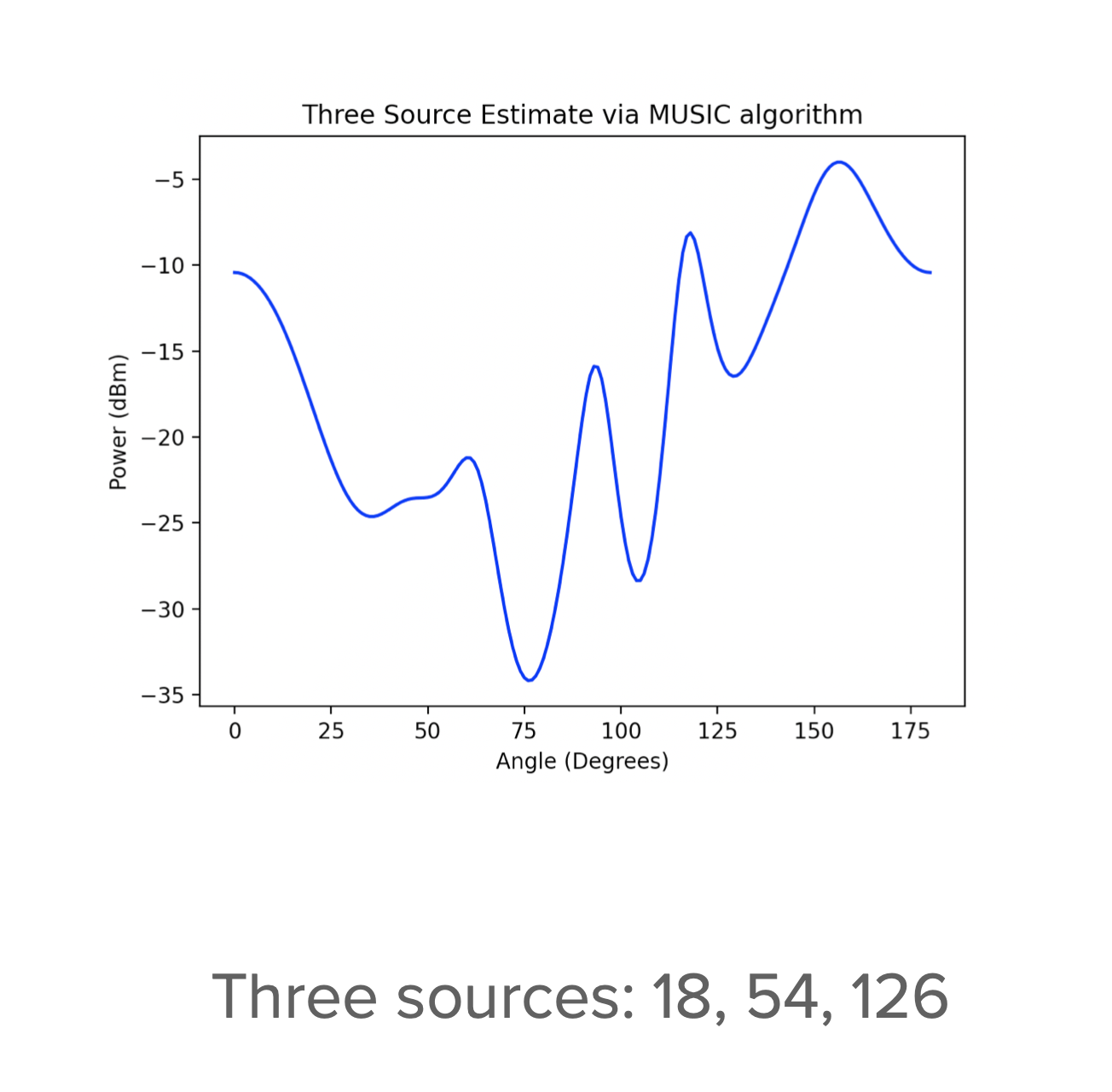}
\label{fig:figure2}
\caption{ 3 Source Simulation Data Plot}
\end{figure}
\begin{figure}[htp]
\includegraphics[width=.45\textwidth]{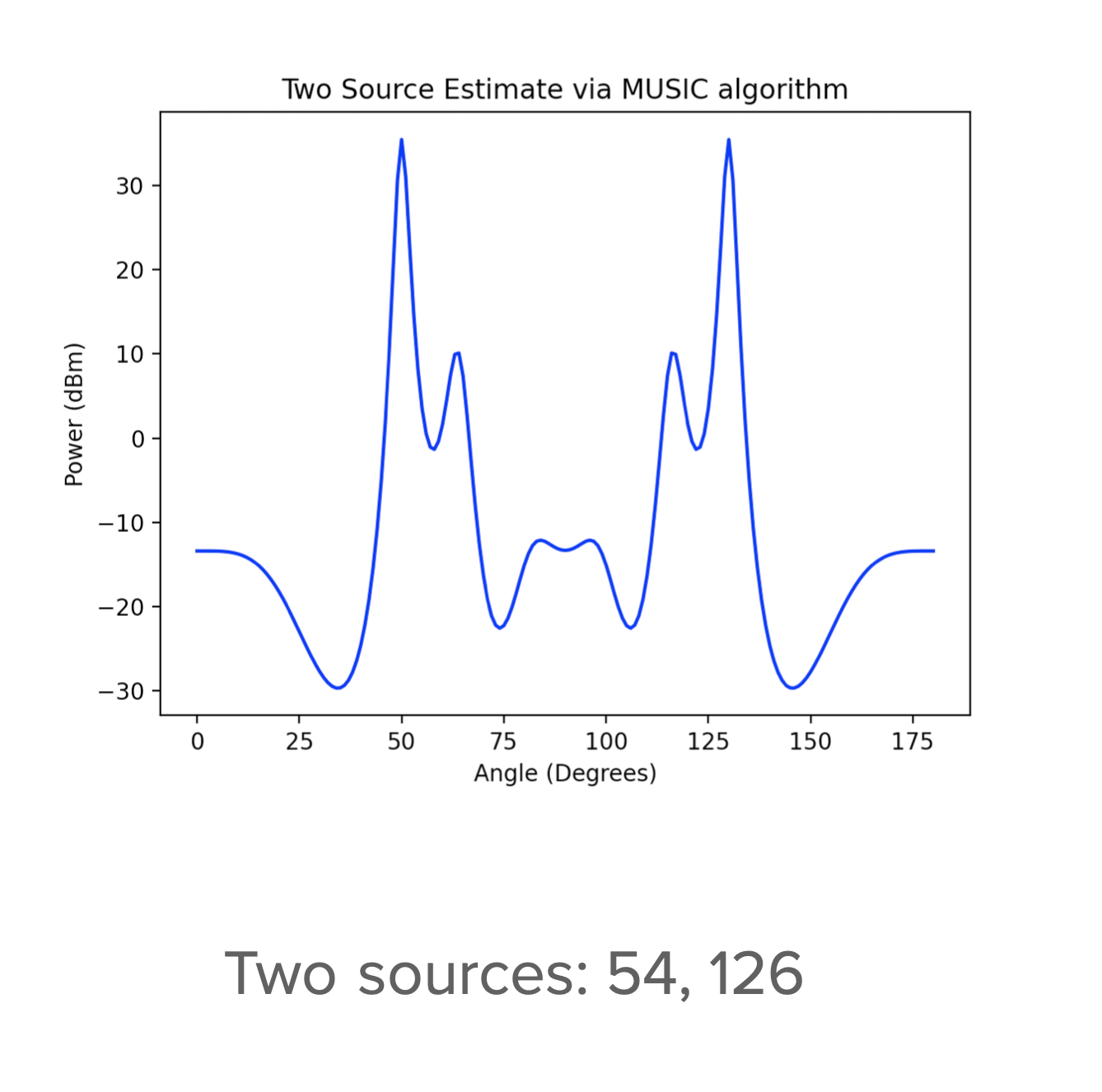}
\label{fig:figure2}
\caption{ 2 Source Simulation Data Plot}
\end{figure}

\begin{figure}[htp]
\includegraphics[width=.45\textwidth]{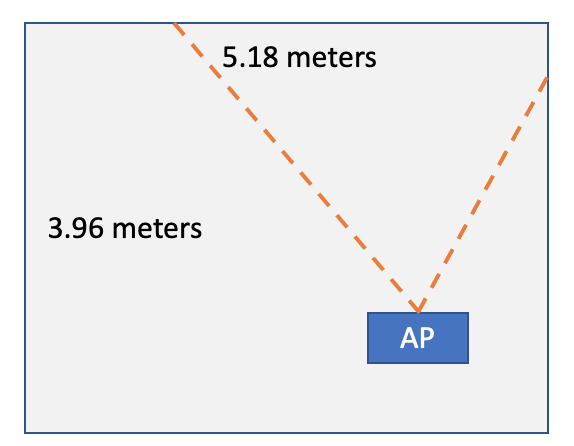}
\label{fig:figure3}
\caption{ This figure illustrates the localization setup that was used for data gathering. The rooms dimensions are listed. Additionally, the AP's location relative to the room is also described. The AP used was the ReSpeaker Mic Array 2.0, which carried an array of 4 microphones positioned in a square formation on the ship. The illustration also shows how the algorithm deployed may project the angle of arrival in the real environment. }
\end{figure}
\begin{figure}[htp]
\includegraphics[width=.45\textwidth]{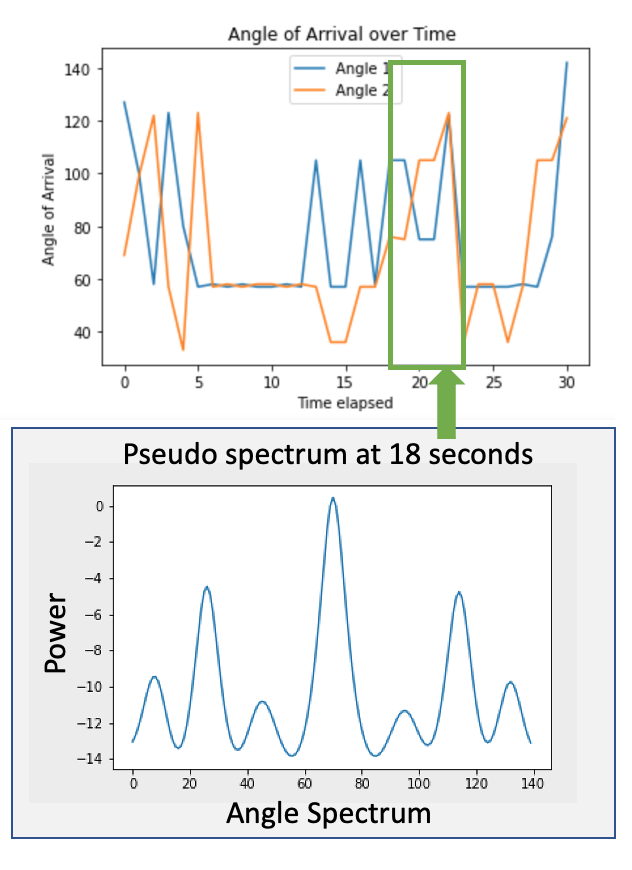}
\label{fig:figure3}
\caption{ This figure illustrates the pseudo spectrum generated by the MUSIC algorithm which went to calculate the angle of arrival for a particular time point. In this case the time point occurs at second 18 (highlighted by the green box). One may notice that the pseudo-spectrum may be having three peaks instead of two. However, due to the scaling of the figure, the difference between the first and third peak looks negligible. The third peak is larger.}
\end{figure}

\begin{figure}[htp]
\includegraphics[width=.45\textwidth]{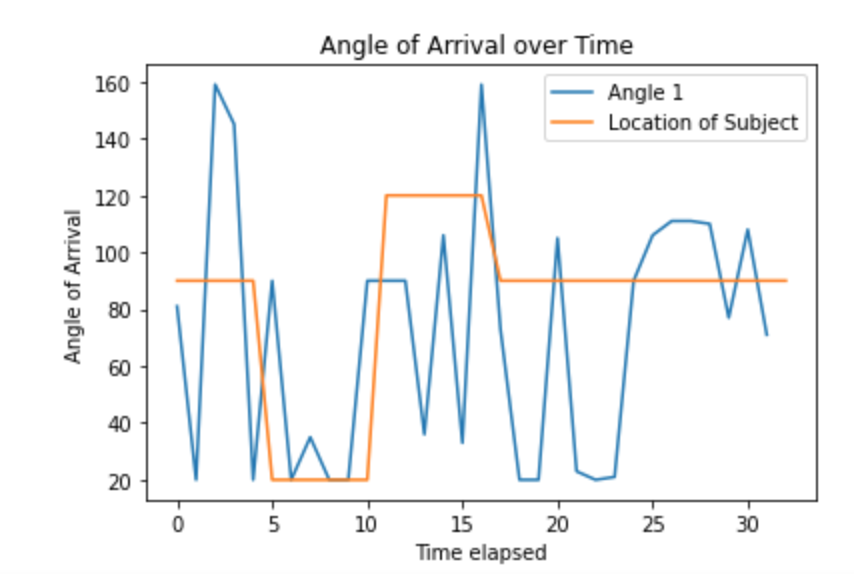}
\label{fig:figure3}
\caption{ This figure illustrates the evaluation of the system that was developed as part of this project. This figure in contrast to the earlier figure is tracking a single source. This is one of the experiments that was carried out. The location of the subject was estimated by manually examining the location of the subject in the video. In this experiment, the subject was clapping and so one can see that the angles estimated are very discrete (as sound was created for a roughly small duration of time). Regardless, the angles still follow the location of the subject relative to the camera. In order to see what activity is going on please see the video in our presentation. }
\end{figure}

\subsection{Physical World Setup}
Our experimental setup uses a ReSpeaker Mic Array 2.0 . It is a microphone array that can connect to a computer using USB and it has four microphones in a two by two grid. We recorded with all four microphone channels at once. We tried a variety of different techniques for recording from this component including using various python modules. However, we ultimately settled on using the Audacity program to record all channels at once and then post processing each microphone's audio into a separate .wav file. This allowed to keep the signals separated for when we applied our MUSIC algorithm's on top of them. 

\begin{figure*}
  \includegraphics[width=\textwidth]{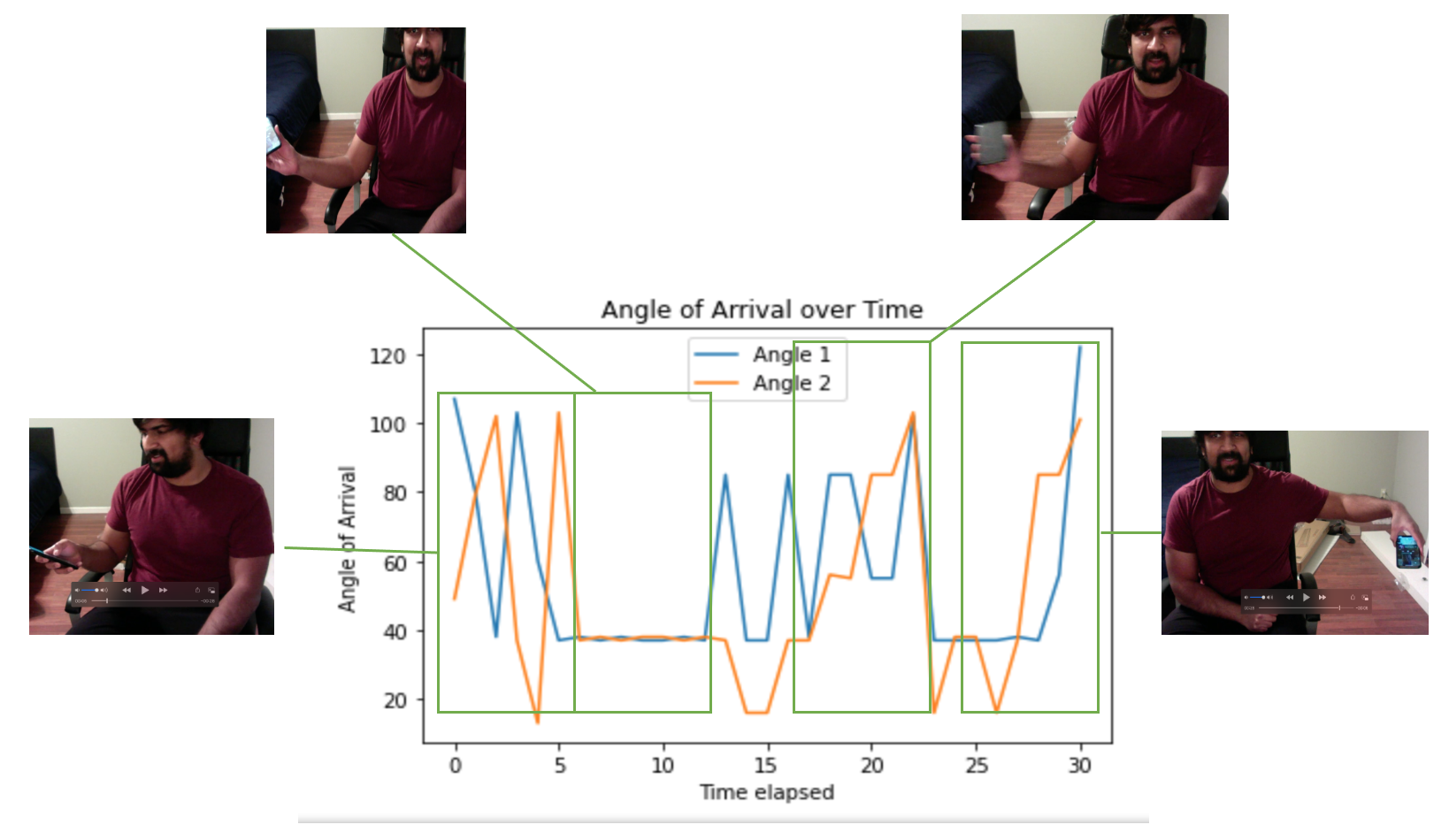}
  \caption{This figure illustrates the use of the multi path MUSIC algorithm for determining the angle of arrival for multiple sources. There are two sources of sound in our scenario: the phone, which is playing audio and the subject themselves, who is talking out loud. The images show what the subject was doing at that particular moment. The graph illustrates the angles of arrival with the highest magnitude determined by the MUSIC algorithm. It is important to note that it only illustrates the two angles at each time step but does not determine if the angle is of the subject or the phone. In the first interval there seems to be a lot of noise, with the subject picking up his phone and orienting himself. In the second interval the subject is talking, but is hard to hear over the sound of the phone. Both the phone and the subject are stationary (hence why it seems to just be predicting the angle of the phone). In the third step, the subject is moving the phone as well as moving his chair. This can be seen by the orange angle shifting from the an angle smaller than the blue to larger. At the last step, the subject quickly moves the phone to the right of him, which is represented by a quick movement in the blue and orange lines and the angle of the blue line points heavily to the right (probably the angle of the phone). }
\end{figure*}

\subsection{Physical World Methodology}
 Each of these input channels has thousands of samples, which correspond to the amplitude and the duration of a single wavelength of sound. Using the input channels, the angle of arrival for each second was calculated. This was done by dividing the total number of samples by the sampling rate, which in this case was 48 kHz. This gave us blocks of samples that were of the size of the sampling rate. Each of these blocks (which were of the size 48,000) was analyzed to determine the angle of arrival of the sound at that location. The number of sources of audio in each interval was set to the number of sources that were used in the demonstration. For example, in our second experiment where we use a phone and a subject, we set the number of audio sources or the number of multi-path for the MUSIC algorithm to 2. We experimented with calculating the number of sources, however the results were not reliable. The output of the MUSIC algorithm is a pseudo-spectrum in which the signal strength of each angle is projected. A series of signal strengths, one for each steering vector corresponding to each theta, is returned. We examined the output of the algorithm and took the angle corresponding to the highest magnitude signal as the angle of arrival. When we had more than 2 sources of audio, we looked at the highest and the second highest power in the MUSIC spectrum instead of looking at the maximum power.

\subsection{Physical World Results}

In order to evaluate the effectiveness of the localization, experiments were done for both the single source audio case and the multi-path audio case. Figure 7 illustrates the experiment that was conducted for the multi-source case. In that experiment, a phone and a subject both talked at the same time. The MUSIC algorithm was used to determine the angles of arrival. The two most prominent angles are plotted in the figure. Another visualization of our system is presented in Figure 5, where we can examine the individual pseudospectrum at a given time step (in this case the 18th sample in the time series). This example shows how the angle of arrival in the multi path setting was derived by looking at the signal with the most power. These figures can give us a qualitative measure of how well the system was working. Figure 6 evaluates how well the MUSIC algorithm and our implementation works in the real world. It shows the location of the subject over time as well as the angle of arrival (single source in this case). We see that that the location of the subject is well tracked. The RMSE measure would be a good way of evaluating the system. However, since there was no baseline system to compare it to, it is not presented here. 

\subsection{Extending Beyond Audio (making it multi modal)}

In order to make the system more precise with regards to localization, we propose to use depth as a tool for gauging where the subject or the source of sound is. Figure 3 can be seen as an illustration of this concept. So far, our system only suggests angles of arrival. However, by using two cameras our system could sense the depth of the object.  Given information on depth and using the $sin$ trigonometric function, it would be possible to evaluate the x and y positions of the sound. 

\section{Leveraging Neural Networks}

\subsection{Event Classification }
After the event detection, a classification model can be used to evaluate if the event was significant or not. This measure is to further decrease the amount of false positive detection of the system. It does this by adding using AI to further filter out positive event detection generated by the averaging of the frames.
The classification system would run after the event detection system runs. This is because polling and running intervals of sound and visual inputs is relatively expensive (as you have to constantly be making passes through the network). This pretrained deep learning model was a Convolutional Neural Network (CNN), trained by researchers at Google \cite{Warden2017Speech}. The network takes in an audio stream and outputs a probability distribution over the tasks it believes took place in the audio. These predefined tasks will then be curated for the project to map to two output classes (Abnormal or Normal). The Normal class was for the normal background sounds normal to systems such as human speech. Abnormal class was for anomalies and tasks seen as non-background. Similar to the audio input processing, a deep learning mode to detect the task being done in a specific video interval can be used but we did not implement in time for the writing of this paper. Predefined tasks would be curated to the same categories as the audio model. The results of the audio and visual models can be used as a baseline off of which multimodal systems can be analyzed from.

\begin{figure}[htp]
\includegraphics[width=.45\textwidth]{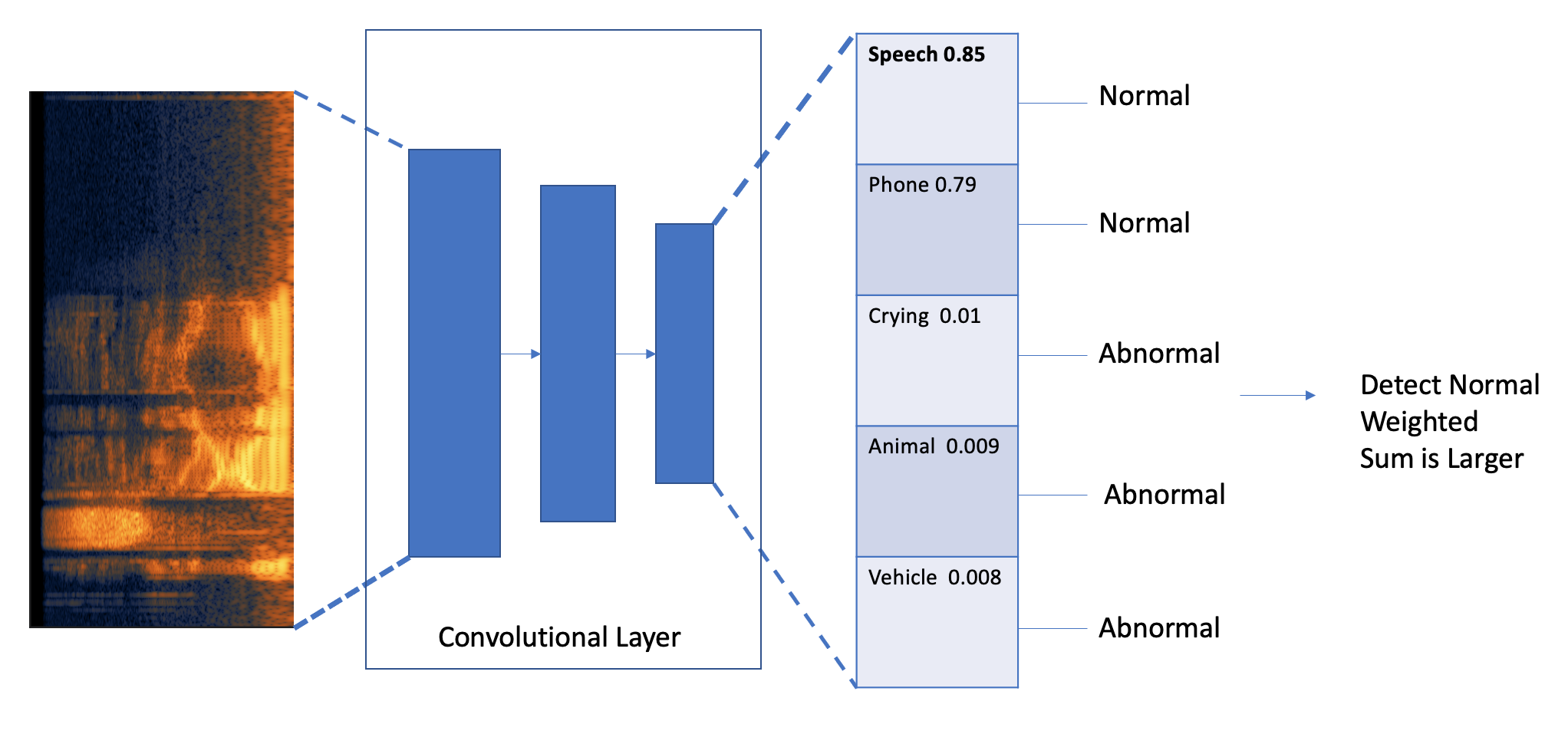}
\label{fig:figure3}
\caption{ This diagram illustrates how adding classification to the audio input, the model is able to be more robust about detecting less false positives in event detection. It is able to be more precise about filtering out data points }
\end{figure}

\subsection{Future Work}
Our work was primarily focused on AoA and understanding the theory and application of it. We were able to get some compelling results which showed that we could track the change of angle of arrival over time in our audio system. Given time, we could make a few key enhancements to make a better system. 
First, we can make our audio processing real time. Instead of recording to audacity, we could go to a lower layer in our computer's firmware and read the individual channels, synchronize them, and pass them through to our computational system. We actually do have most of this working already but the problem is we couldn't figure out how to split out the channels in real time so we couldn't put the data in a form our algorithms could understand.
Second, we could take the angle of arrival calculations in real time and project onto the screen a visualization of where the audio is coming from. This would help us combine our audio system with our visual system to create more informed decisions on intrusion detection. It would also be further visual confirmation that our AoA calculations have merit. We did compare our results to ground truth values in our simulation exercise but it is more pleasing to visually see the result projected onto real world data that we gathered.
Third, we can combine our audio-visual system with the ML model we experimented with to not only detect an event but also classify it. A classification of an audio event combined with a detection of an event in the multi modal system can further inform a security system on what to do and whether or not to alert the user.

\subsection{Challenges}
Our primary challenge stemmed from the fact that we wanted to do a project that had a hardware component. We wanted to do our own data gathering and that required research into various hardware components. Our initial attempts failed as the Arduino cam and board did not provide sufficient data throughput for our visual system. Our audio system ran into a different set of issues as we needed to find a way to get an array of microphones to record simultaneously. Eventually, we found the ReSpeaker hardware component that allowed us to do that simply. It was an interesting challenge figuring out how to sample from the hardware component properly. We had control over variables such as the sample rate and it demonstrated that there was more to consider than just pressing record when dealing with signal gathering in the field.

\subsection{Conclusions}
Overall, the project was a good learning experience in primarily implementing an algorithm to detect Angle of Arrival of multiple signals in a real world scenario. We were able to implement our own Music algorithm on simulation data for multiple sources. We were also able to perform angle of arrival calculation over time for our real life gathered data. The project opened us up to a variety of techniques and challenges in signal based research and was rewarding.

\bibliographystyle{acl_natbib}
\bibliography{sample-bibliography}

\end{document}